# Responses of the Neurobiological Craving Signature to smoking versus alternative social rewards predict craving and monthly smoking in adolescents

Maddalena Tamellini[1], Joyce Dieleman[2,3,4], Guillaume Sescousse[1], Maartje Luijten[3], & Leonie Koban[1]*

[1] Lyon Neuroscience Research Center (CRNL), INSERM U1028, CNRS UMR5292, Claude Bernard Lyon 1 University, Lyon, France.
[2] Trimbos Institute, Utrecht, The Netherlands.
[3] Behavioral Science Institute, Radboud University, Nijmegen, The Netherlands.
[4] GGD Noord- en Oost Gelderland, Warnsveld, The Netherlands.

*Please address correspondence to:

Dr. Leonie Koban
Centre de Recherche en Neurosciences de Lyon
Inserm U1028 - CNRS UMR5292 - Université Claude Bernard Lyon1
Institut des Épilepsies IDEE
59 Boulevard Pinel
69500 BRON
Email: leonie.koban@cnrs.fr
Telephone: +33 (0)4 81 10 65 93

**Abstract**

Smoking remains the leading cause of preventable mortality worldwide. Adolescents are particularly vulnerable to the development of tobacco addiction due to ongoing brain maturation and susceptibility to social influences, such as exposure to environmental tobacco smoke (ETS). Craving – the strong desire to use drugs – already emerges with non-daily tobacco use and predicts continued use and relapse. However, the roles of craving and ETS exposure during the early stages of tobacco use in adolescence remain poorly understood. In this pre-registered study, we harness a recently developed fMRI marker of craving – the *Neurobiological Craving Signature* (NCS) – to compare craving-related brain responses to smoking versus social cues in adolescent Experimental Smokers (N=100) and Non-smokers (N=48) with varying levels of ETS exposure levels. Results showed that NCS responses to smoking cues compared to alternative social rewards were higher in Experimental Smokers compared to Non-smokers and predicted individual differences in self-reported craving and monthly smoking. Both smoking behavior and NCS responses were correlated with the relative amount of ETS exposure from peers compared to exposure from family members. Together, these findings indicate a heightened sensitivity of craving-related brain circuits already during experimental smoking and highlight the important role of peer social norms on craving and smoking initiation in the critical period of adolescence.

## Introduction

Tobacco smoking causes more than 8 million deaths each year, including an estimated 1.3 million non-smokers who are exposed to second-hand smoke (WHO, 2023). This makes smoking a major cause of preventable deaths worldwide and a significant public health burden. Adolescents and young adults are at particularly high risk of developing tobacco addiction due to ongoing brain maturation (Casey et al., 2008, Baker et al., 2015) and a higher sensitivity to social influences (Ciranka & Van Der Bos, 2019). Craving for tobacco is the most reported symptom appearing with minimal nondaily tobacco use (DiFranza & Wellman, 2005). Thus, characterization of craving and its underlying brain mechanisms may serve as an early predictor of smoking uptake. Here, we harness the *Neurobiological Craving Signature* (NCS) – a recently developed fMRI-based neuromarker of drug craving (Koban, Wager & Kober, 2023) – in a large sample of adolescents (N=148) to test the role of craving-related brain responses and social context in adolescent experimental smokers compared to non-smokers.

The NCS predicts within-person variability of drug and food craving based on distributed brain activation patterns and discriminates between substance users and controls with over 80% out-of-sample accuracy (Koban, Wager & Kober, 2023), making it one of the first candidate neuromarkers in addiction. The NCS has been developed in a sample of adults, most of whom with severe substance use disorder or tobacco addiction. It therefore remains unknown whether the NCS is sensitive to craving and smoking in adolescents who are in the critical initial phase of smoking uptake and at the risk of becoming regular smokers. We address this open question in a sample of non-smoking and experimental smoking teenagers and test whether NCS responses to smoking versus non-smoking-related cues predict between-person differences in self-reported craving, smoking behavior, and environmental tobacco smoke exposure (ETS) – a social-contextual factors known to influence smoking (Miller et al., 2009; Okoli et al., 2016).

Craving – defined as a persistent and intense desire to use drugs – is a core symptom of substance use disorders, including tobacco addiction (O'Brien et al., 1998; Sayette 2016; Vafaie & Kober, 2022, Gauld et al., 2023). Craving is considered one of the most central symptoms of addiction and a significant predictor of drug use and relapse (Allenby et al., 2020; Chirokoff et al., 2023; Vafaie & Kober, 2022). Craving for tobacco is also the first and most commonly reported symptom emerging even with minimal nondaily tobacco use (DiFranza et al., 2000; Doubeni et al., 2010). In experimental smoking teenagers, craving appears early and, once present, leads to a higher probability of attaining daily smoking and conversion to tobacco addiction as defined by ICD-10 (Gervais et al., 2006). Thus, behavioral and neurophysiological measures of craving, such as the NCS, may serve as early predictors of smoking initiation, continuation, and development of tobacco addiction. In the long term, identifying reliable biomarkers of vulnerability to smoking, especially among youth, may contribute to (1) identify at-risk individuals, (2) assess the effects of social and environmental factors (e.g., stress, ETS) on smoking uptake, and (3) provide biological targets to evaluate the effectiveness of smoking cessation treatments.

Adolescence is a critical period for developing tobacco addiction. The majority of adult smokers began smoking during adolescence, with an earlier age of onset being associated with higher risk of dependence later in life (Kendler et al., 2013; Reitsma et al., 2021). This vulnerability might be due to the ongoing maturation of brain areas involved in reward and inhibitory control, making adolescents more susceptible to risky behaviors, such as smoking (Casey et al., 2008; Lydon et al., 2014; Castro, Lotfipour & Leslie, 2023). Adolescents are also very susceptible to peer influence and perceived social norms, and certain social and environmental factors may further increase the vulnerability to nicotine use and the development of addiction (Simons-Morton & Farhat, 2010; Piña et al., 2018).

Among these, exposure to environmental tobacco smoke (ETS) from family members and peers, is thought to be especially important. ETS exposure has not only been linked to various adverse health outcomes, including higher risks of cancer, ischemic heart disease, stroke, and type-2 diabetes (Flor et al., 2024), but could also affect smoking-related behaviors and health in several ways. As suggested by both rodent and human studies, one possibility is that ETS may have a direct physiological addictive effect due to the nicotine found in passive smoke (Small et al., 2010; Brody et al., 2011). Indeed, the nicotine concentration in passive smoke has been compared to that absorbed during non-daily smoking (Al-Delaimy et al., 2001). Children and adolescents could be particularly vulnerable to the effects of ETS exposure (Schuck et al., 2013; Bélanger et al., 2008) because of their capacity to absorb higher quantities of nicotine than adults (Willers et al., 1995).

In addition to these physiological pathways, ETS exposure also conveys important social information about smoking. Other people's thoughts, expectations and behaviors are key determinants of health-related behaviors, including drug use (Pelloux et al., 2019), especially during adolescence and early adulthood (Knoll et al., 2015; Gardner & Steinberg, 2005; Somerville et al., 2013). Thus, beyond its direct pharmacological impact, ETS exposure may signal smoking-related norms within an adolescent's social environment, especially among friends. While both parental and peer influences can affect smoking-related behaviors, peer norms often gain increasing prominence during adolescence (Knoll et al., 2015). This perspective may help explain why ETS exposure from family and friends has been linked to an increased risk of smoking initiation and a greater likelihood of relapse after cessation attempts in both adolescents and adults (Racicot et al., 2011; Leonardi-Bee et al., 2010; Okoli & Kodet, 2015; Okoli et al., 2016). However, the brain mechanisms through which social influences during adolescence shape tobacco-related symptoms such as craving remain largely unknown.

Experimentally, craving for cigarettes can be triggered by various cues (e.g. seeing someone else smoking or smelling cigarette smoke). Nicotine-dependent individuals typically show heightened physiological arousal, craving, and smoking behavior compared to non-smokers in response to such stimuli (Carter & Tiffany, 1999; Yalachkov, Kaiser & Naumer, 2012). In this study, we employed a smoking cue-reactivity task featuring visual cues (photos) of social smoking scenes, neutral social interactions, and romantic social interactions – a powerful alternative reward, especially in a population of adolescents (Fisher et al., 2016; Xu et al., 2012). By including socially embedded smoking cues, this paradigm reflected situations in which ETS exposure commonly occurs and allowed us to assess how it may trigger cigarette craving in non-smoking and experimental smoking adolescents.

Meta-analyses of human fMRI studies have identified several brain regions involved in craving and drug cue reactivity (Jasinska et al., 2014; Kühn & Gallinat, 2011; Noori et al., 2016; Owens et al., 2018; Kober et al., 2010; Rubinstein et al., 2011; Engelmann et al., 2012; Goldstein & Volkow, 2002; Murphy et al., 2018). Key regions within the mesocorticolimbic system, such as the ventral striatum, amygdala, anterior cingulate cortex, prefrontal cortex and insula, play a central role due to their involvement in reward, motivation, and goal-directed behavior. The hippocampus and orbitofrontal cortex contribute to drug-related memory and impaired inhibitory control (Jasinska et al., 2014; Rubinstein et al., 2011). Finally, the nigrostriatal system, encompassing the substantia nigra, caudate, putamen, and globus pallidus, facilitates habit learning and transition from controlled to automatic behavior (Jasinska et al., 2014; Noori et al., 2016; Engelmann et al., 2012).

Although numerous studies have explored the neural basis of craving in adult smokers and nonsmokers, there is very little research in humans on neurobiological changes arising from first uses of cigarettes and exposure to environmental tobacco smoke during adolescence. Only a few neuroimaging studies investigated whether non-daily smoking and ETS exposure could affect brain functioning related to motivational processes and inhibitory control in non-smoking and experimental smoking adolescents (Boormans et al., 2021; Dieleman et al., 2020; Dieleman et al., 2022). The current pre-registered study reanalyzed two longitudinal datasets, including unpublished data from 100 experimental smoking adolescents and previously published data from 48 non-smokers (Boormans et al., 2021), to investigate whether non-daily smoking and ETS exposure influence adolescents' cue-induced craving and craving-related brain activity, as measured by the NCS.

We hypothesized that social smoking-related versus neutral and romantic interactions cues would be associated with higher NCS responses in experimental smokers, compared to never-smoking adolescents, while both smokers and never-smokers may show increased NCS responses to romantic compared to neutral cues. Second, we hypothesized that both non-daily-smoking and ETS exposure increase craving and craving-related brain activity even in adolescents without established smoking patterns. Thus, higher levels of current monthly smoking, ETS exposure, cotinine (a proxy for nicotine absorption) levels and self-reported craving would be associated with higher NCS responses to Smoking compared to Neutral and Romantic cues. Finally, we tested whether individual differences in NCS responses to Smoking (compared to Neutral and Romantic) cues at baseline were associated with smoking frequency and behavior at six-months and at one-year follow-up in the group of experimental smoking adolescents.

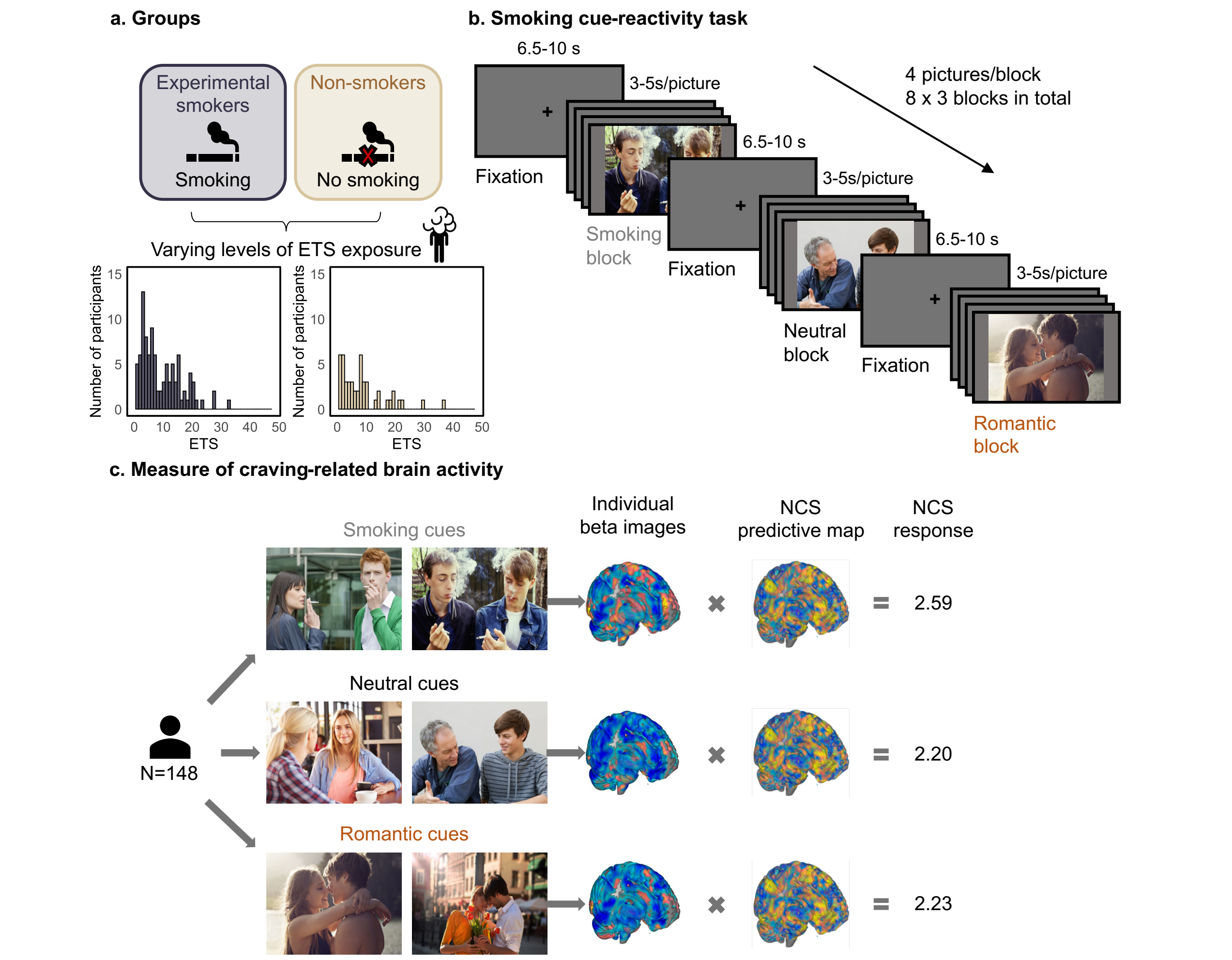

Fig. 1: **Experimental design and analysis approach**. **a.** Both groups (Experimental Smokers and Non-Smokers) were exposed to varying levels of environmental tobacco smoke (ETS) in their social environment. **b.** The smoking cue-reactivity task consisted of 24 consecutive blocks containing 4 pictures each. In each block, participants saw photographs of social cues (always featuring two or more people) from one of three different experimental conditions: 8 blocks of Smoking interactions, 8 blocks of non-smoking (Neutral) interactions, and 8 blocks of Romantic interactions (which constitute an alternative reward). **c.** The NCS weight map specifies each voxel's positive or negative contribution to the prediction of craving. The NCS was applied to individual first-level beta images for the contrasts [Smoking>Baseline], [Neutral>Baseline] and [Romantic>Baseline]. This allowed to compute one score per cue-type and per individual, reflecting the average craving-related brain activity associated with each cue for each participant.

## Results

### *Demographics and smoking behavior*

This study compared two participant groups (Table 1). Non-smokers included a total of 48 participants who never smoked a complete cigarette in their life. Experimental smokers included a total of 100 participants who smoked between 5 and 500 cigarettes in their lifetime. As detailed in Table 1, Non-smoking and Experimental smoking adolescents did not differ in terms of gender, age and pubertal development. There was no significant difference in ETS exposure levels, familial risk to develop nicotine dependence – measured with a composite score assessing smoking habits of participants' parents and grandparents – and maternal smoking during pregnancy between the two groups. As expected, Experimental smoking adolescents smoked more cigarettes (p<0.001) but also used more alcohol (p=0.003) and cannabis (p<0.001), had higher salivary cotinine levels (p<0.001), and reported higher craving for cigarettes (p<0.001) than non-smoking adolescents. Compared to Non-smokers, Experimental smokers also scored higher on scales indexing loss of control over tobacco use, like the HONC (p<0.001) and the AOTS (p<0.001). However, these scores were generally below the conventional limit of high dependence in both groups, with just a few smokers having above-threshold scores (see Supplementary Fig. S1). Experimental smokers also had higher scores for the smoking-related self-efficacy questionnaire (p<0.001), which assessed their temptation to smoke when facing both internal and external stimuli, showing that they were less confident in their ability to refrain from smoking than Non-smokers. As confirmed by the similar scores in the Smoking-Specific Parenting questionnaire (p=0.866), the parenting practices related to smoking behavior did not differ significantly between the two groups.

**Table 1. Comparison of the characteristics of the two experimental groups.**

| *Variables* | *Experimental smokers* | *Nonsmokers* | *N Experimental-smokers* | *N Non-smokers* | *U / χ2 / V* | *p* |
|---|---|---|---|---|---|---|
| Gender (%) | 30% boys<br>70% girls | 27% boys<br>73% girls | 100 | 48 | $2.97 \times 10^{-02}$ | 0.863 |
| Age (years) | 16.50 ± 1.13 | 16.56 ± 1.09 | 100 | 48 | 2315 | 0.718 |
| Pubertal development status | 3.51 ± 0.43 | 3.57 ± 0.37 | 100 | 48 | 2264 | 0.573 |
| Familial risk | 1.99 ± 2.35 | 2.91 ± 2.93 | 91 | 46 | 1703 | 0.071 |
| Mothers smoked during pregnancy *% yes* | 13% yes | 22% yes | 91 | 46 | 1.08 | 0.298 |
| **AUDIT** ** | 8.68 ± 4.76 | 6.38 ± 5.09 | 96 | 39 | 2493.5 | 0.003 |
| **Cannabis life time use** *** | 16.44 ± 34.71 | 1.06 ± 4.56 | 100 | 48 | 3649 | $3.07 \times 10^{-8}$ |
| **Saliva cotinine (μg/L)** *** | 6.56 ± 15.49 | 1.11 ± 5.17 | 100 | 48 | 3332.5 | $7.58 \times 10^{-5}$ |
| ETS exposure | 9.03 ± 6.91 | 7.98 ± 773 | 100 | 48 | 2723.5 | 0.185 |
| **Monthly smoking** (cigarettes/month) *** | 7.22 ± 9.46 | 0.00 ± 0.00 | 100 | 48 | 3240 | $7.46 \times 10^{-15}$ |
| **Self-reported craving** *** | 16.91 ± 20.64 | 0.12 ± 0.73 | 100 | 48 | 4114.5 | $6.74 \times 10^{-14}$ |
| FTND | 3.07 ± 0.33 | 3.00 ± 0.00 | 100 | 11 | | |
| **HONC** *** | 1.40 ± 1.73 | 0.10 ± 0.37 | 100 | 48 | 3664 | $7.57 \times 10^{-9}$ |
| **AOTS** *** | 3.13 ± 2.88 | 0.15 ± 0.41 | 100 | 48 | 4450.5 | $5.70 \times 10^{-18}$ |
| QSU | 16.41 ± 11.41 | 21.09 ± 24.19 | 100 | 11 | | |
| **SSEQ** *** | 27.53 ± 8.42 | 13.96 ± 7.07 | 100 | 48 | 4533 | $1.39 \times 10^{-18}$ |
| SSP | 48.90 ± 8.63 | 49.27 ± 7.08 | 100 | 48 | 2358.5 | 0.866 |

*Note:* Continuous variables (age, pubertal status, familial risk, alcohol and cannabis use, cotinine levels, ETS exposure, monthly smoking, self-reported craving and questionnaires scores) are reported as mean ± standard deviation. Dichotomous variables (gender and smoking during pregnancy) are reported as the percentage of boys and girls, and the percentage of participants whose mothers smoked during pregnancy. Variables for which values significantly differed between the two groups are indicated in bold (two-sided Mann-Whitney U-tests for continuous variables, one-sample Wilcoxon test for continuous variables with a variance of 0, Chi-squared tests for dichotomous variables, *p<0.05, **p<0.01, ***p<0.001). Only 11 out of 48 Non-smokers answered the FTND and the QSU, as they had previously smoked a drag but never a full cigarette, which was the exclusion criteria for this group. Given this small subgroup and the resulting limited statistical power, we did not conduct group comparisons for these questionnaires. AUDIT, Alcohol Use Disorders Test; FTND, Fagerström Test for Nicotine Dependence; HONC, Hooked On Nicotine Checklist; AOTS, Autonomy Over Tobacco Scale; QSU, Questionnaire for Smoking Urges; SSEQ, Smoking-related Self-Efficacy; SSP, Smoking Specific Parenting.

Experimental smokers' questionnaire scores did not vary substantially across the three different measurement time points (Supplementary Table S1), except for monthly smoking and the Hooked on Nicotine Checklist (HONC). Post-hoc comparisons for monthly smoking revealed a trend toward higher scores at one-year follow-up compared to both baseline (mean difference [MD]=5.97, SE=2.49, t(156)=2.40, p=0.053) and six-month follow-up (MD=5.73, SE=2.49, t(156)=2.30, p=0.068). No significant difference was observed between baseline and six-month follow-up (MD=0.24, SE=2.49, t(156)=0.096, p=1.00). For the HONC, participants reported greater symptoms of nicotine dependence at both six-months (MD=0.71, SE= 0.19, t(198)= 3.72, p=0.0008) and one-year (MD=0.55, SE=0.19, t(198)=2.88, p=0.013) follow-ups compared to baseline. Scores did not differ significantly between six-months and one-year follow-ups (MD=-0.16, SE=0.19, t(198)=-0.84, p=1.00).

***NCS differences between Experimental smokers and Non-smokers***

Craving-related brain activity during the smoking cue-reactivity task was assessed in both groups by computing responses (mean ± sem) of the Neurobiological Craving Signature (NCS) to Smoking (Non-smokers: 1.97±0.08, Experimental smokers: 2.08±0.06), Neutral (Non-smokers: 2.04±0.09, Experimental smokers: 2.04±0.07) and Romantic cues (Non-smokers: 2.26±0.08, Experimental smokers: 2.22±0.06; Fig. 2a). Higher NCS scores indicate more activity in distributed craving-related brain areas and higher levels of brain-predicted craving. A mixed-effects ANOVA revealed a main effect of the cue type (F(2,438)=7.11, p=0.001), with planned t-tests (Bonferroni corrected) showing that, across both groups, Romantic cues elicited higher NCS responses than Smoking (df=147, t=5.13, p<0.001) and Neutral cues (df=147, t=4.52, p<0.001), whereas Smoking and Neutral cues elicited similar NCS responses (df=147, t=0.18, p=1.00). There was no main effect of group (F(1,438)=0.003, p=0.95) on NCS responses to social cues. The interaction between group and cue-type showed a trend (F(2,438)=2.21, p=0.11).

To explore this trend and test our hypothesis that Experimental smokers would show increased craving-related brain activity to smoking cues relative to alternative social cues compared to Non-smokers, we computed NCS difference scores to Smoking versus Neutral (Fig. 2b) and to Smoking versus Romantic cues (Fig. 2c). NCS responses to Smoking>Neutral cues showed a trend towards a group difference in the expected direction, with Experimental smokers having slightly increased craving-related brain activity compared to Non-smokers (U=2748, p=0.077). Moreover, compared to an alternative social reward (Romantic cues), Smoking cues elicited higher NCS responses, thus a relatively stronger craving-related brain activity, in Experimental smokers than in Non-smokers (U=2832, p=0.039).

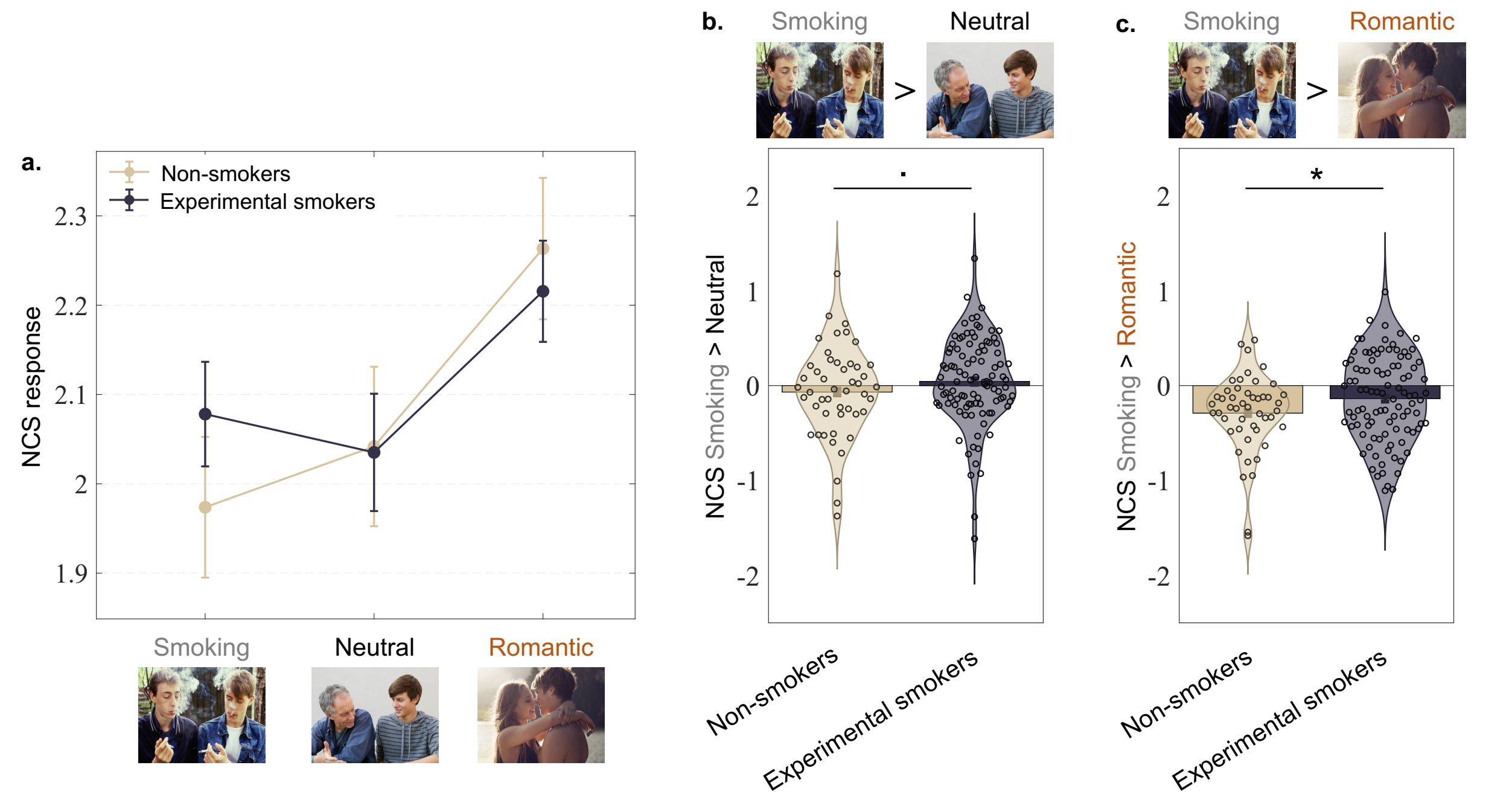

**Fig. 2: Group differences in NCS responses during the smoking-cue reactivity task. a.** Non-smoking (N=48) and Experimental smoking (N=100) adolescents NCS responses to Smoking, Neutral and Romantic social cues. Main effect of cue-type (within-subjects factor), group (between-subjects factor) and their interaction were assessed with a mixed-effect ANOVA and post hoc comparisons with Bonferroni correction. **b.** NCS responses to [Smoking > Neutral] cues for Non-smokers (mean±sem=-0.07±0.07) and Experimental smokers (mean±sem=0.04±0.05). Group differences were assessed with a one-tailed Mann Whitney U test (. p<0.1). **c.** NCS responses to [Smoking > Romantic] cues for Non-smokers (mean±sem=-0.29±0.06) and Experimental smokers (mean±sem=-0.14±0.05). Group differences were assessed with a one-tailed Mann Whitney U test (* p<0.05).

### *Associations with current salivary cotinine levels, ETS exposure, monthly smoking and self-reported craving*

As shown in Fig. 2, NCS responses showed substantial variability between participants and especially within the group of Experimental smokers, which may reflect the different degrees of smoking among these participants and their social environment. We therefore investigated individual differences in monthly smoking, NCS responses, self-reported craving, salivary cotinine and ETS exposure (Fig. 3a).

First, we confirmed that greater monthly cigarette consumption was linked to higher self-reported craving levels ($\tau$=0.53, p<0.001) and that higher salivary cotinine levels were associated with increased ETS exposure ($\tau$=0.20, p=0.001), monthly smoking ($\tau$=0.40, p<0.001) and self-reported craving ($\tau$=0.21, p=0.001). However, in contrast to our initial hypothesis, greater overall ETS exposure was not significantly related to monthly smoking ($\tau$=0.10, p=0.110) and self-reported craving ($\tau$=0.08, p=0.215). This pattern may reflect the recruitment strategy, which was designed to include participants with a broad range of ETS exposure across both groups. Consequently, Experimental smoking and Non-smoking adolescents experienced comparable overall ETS levels (see Table 1), potentially limiting the ability to detect associations between passive nicotine exposure and smoking behavior and craving.

One potential pathway through which ETS exposure might influence smoking behavior is passive nicotine intake, which may sensitize brain reward pathways similarly to nicotine in regular smokers. An alternative, non-exclusive hypothesis is that ETS may reflect social norms about smoking in the participant's environment. Social norms shape behavior, particularly in adolescence and early adulthood, where peer influence is a key determinant of health-related behaviors, whereas parents may become less influential or even targets of oppositional behavior. The overall measure of ETS exposure combined second-hand smoke exposure from parents, other family members, babysitters, friends and others. To explore the role of social influence in more detail, we conducted a non-preregistered analysis distinguishing ETS exposure from friends and family, as well as their relative contribution (ETS friends – family), and tested associations with other variables of interest. In line with the idea that adolescents were more influenced by smoking peers than by smoking family members, second-hand exposure from friends was positively correlated with monthly smoking ($\tau=0.29$, $p<0.001$) and craving ($\tau=0.21$, $p<0.001$), whereas overall ETS and ETS from family were not. This association was also significant for the relative score ETS friends – ETS family ($\tau=0.20$, $p=0.001$ and $\tau=0.12$, $p=0.046$).

Second, we tested the association with craving-related brain responses. In line with our hypothesis, higher NCS responses to both Smoking>Neutral ($\tau=0.14$, $p=0.024$) and Smoking>Romantic cues ($\tau=0.14$, $p=0.020$) were significantly associated with greater self-reported craving for cigarettes (Fig. 3b, 3c).
Similarly, higher cigarette consumption was associated with greater NCS responses to Smoking relative to Romantic ($\tau=0.13$, $p=0.026$; Fig. 3e), but not Neutral ($\tau=0.05$, $p=0.396$; Fig. 3d), cues. Most interestingly, NCS responses were significantly and positively associated with relative exposure to ETS from peers versus family members ([Smoking>Neutral]: $\tau=0.12$, $p=0.030$; [Smoking>Romantic]: $\tau=0.12$, $p=0.039$). This confirms that individuals who experience more ETS exposure from friends than family show the highest craving-related brain response when presented with smoking-related cues compared to non-smoking related cues (Fig. 3f, 3g).

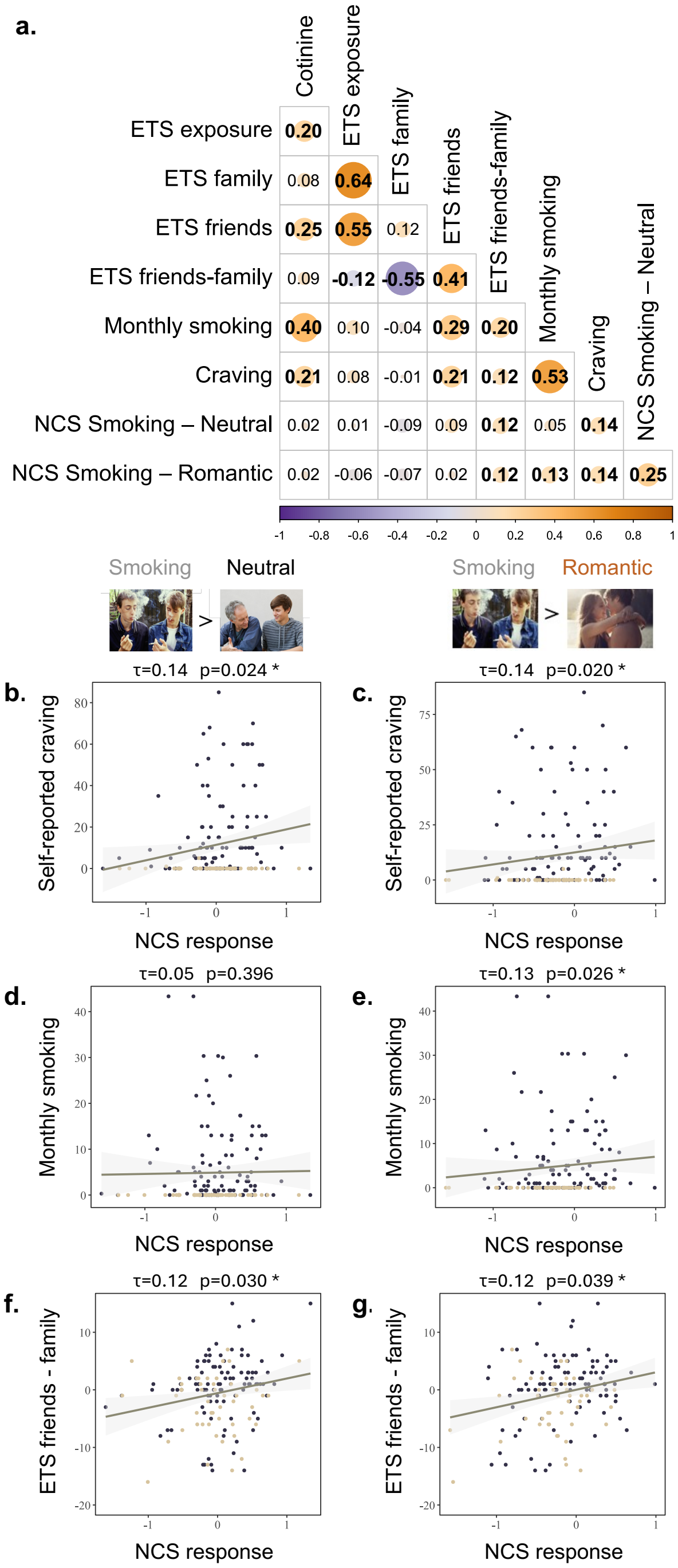

| | Cotinine | ETS exposure | ETS family | ETS friends | ETS friends-family | Monthly smoking | Craving | NCS Smoking – Neutral |
|---|---|---|---|---|---|---|---|---|
| ETS exposure | **0.20** | | | | | | | |
| ETS family | 0.08 | **0.64** | | | | | | |
| ETS friends | **0.25** | **0.55** | 0.12 | | | | | |
| ETS friends-family | 0.09 | **-0.12** | **-0.55** | **0.41** | | | | |
| Monthly smoking | **0.40** | 0.10 | -0.04 | **0.29** | **0.20** | | | |
| Craving | **0.21** | 0.08 | -0.01 | **0.21** | **0.12** | **0.53** | | |
| NCS Smoking – Neutral | 0.02 | 0.01 | -0.09 | 0.09 | **0.12** | 0.05 | **0.14** | |
| NCS Smoking – Romantic | 0.02 | -0.06 | -0.07 | 0.02 | **0.12** | **0.13** | **0.14** | **0.25** |

**Fig. 3: Associations between the NCS responses to Smoking versus Neutral and Romantic cues and smoking-related variables. a.** Correlation matrix between NCS responses to [Smoking>Neutral] cues, [Smoking>Romantic] cues, salivary cotinine levels (µg/L), ETS exposure (overall, from family, from friends and friends-family), monthly smoking (cigarettes/month) and self-reported craving. Kendall's tau correlations were used to test these associations. Significant correlations are reported in bold. Correlations between NCS responses to Smoking>Neutral and Smoking>Romantic cues and self-reported craving (**b, c**), monthly smoking (**d, e**) and the relative ETS exposure from friends versus family **(f, g**). Non-smokers (n=48) are indicated in beige and Experimental smokers (n=100) are indicated in dark blue. In gray, the regression line computed across all participants (n=148), and its 95% confidence interval.

### *Individual differences in NCS responses and smoking-related behavior at follow-up*

Lastly, we assessed whether individual differences in NCS responses to Smoking compared to Neutral cues at baseline were associated with Experimental smokers' future smoking-related behavior (Supplementary Table S4).

There was no significant association between baseline NCS responses to Smoking>Neutral cues and follow-up smoking frequency and self-reported craving. NCS responses to Smoking>Romantic cues were not significantly correlated with follow-up self-reported craving or monthly smoking. We note that the limited variability in smoking-related behaviors observed at baseline, as well as the very small changes in craving and smoking behavior between baseline and follow-up (Supplementary Table S1) might have limited the possibility to detect such associations.

## Discussion

Most smokers start experimenting with smoking during adolescence. Understanding the initial stages of tobacco addiction during this critical period is thus crucial for reducing the exposure to factors facilitating smoking uptake and for preventing continued tobacco use. This study investigated craving-related brain responses to smoking cues in adolescents experimenting with smoking versus adolescent non-smokers, and their link to environmental tobacco smoke (ETS) exposure from different sources (peers versus family members). For this purpose, we harnessed an fMRI-neuromarker of drug and food craving—the recently and independently developed *Neurobiological Craving Signature* (NCS, Koban et al., 2023) and compared its responses to social smoking cues, neutral, and romantic social scenes.

Overall, NCS responses in both groups of adolescents were highest for romantic social cues – a powerful alternative social reward. NCS responses to Smoking versus Romantic cues were also higher in Experimental smokers than Non-smokers and were associated with monthly smoking at baseline. This suggests that craving-related brain responses to smoking cues are already increased relative to alternative social rewards at an early stage of smoking, reflecting an overvaluation of drug rewards at the expense of alternative, social rewards. Further and in line with our predictions, NCS responses to Smoking versus Neutral and versus Romantic cues predicted self-reported craving of participants. While we did not find significant associations between NCS responses, cotinine levels, and overall current ETS exposure, NCS responses as well as smoking behavior were significantly associated with the relative ETS exposure from friends versus family members, suggesting that normative social influence from peers might be more important for craving than ETS exposure from family or overall exposure, at least during this specific phase of smoking initiation and developmental period, adolescence.

The NCS was originally trained to predict within-subject differences in craving across several types of drug and food cues in several cohort of adults, most of them individuals with diagnosed substance use disorder or longstanding nicotine addiction (Koban et al., 2023). The present results show that increased NCS responses significantly predict higher self-reported craving for smoking in adolescents. This is noteworthy for several reasons. First, it provides novel evidence that the NCS also captures between-subject differences in craving. Second, it shows that the NCS generalizes to a different population (adolescents in the Netherlands), and different experimental setting and stimuli sets, reinforcing its validity as a neural marker of craving. Third, it demonstrates the value of the NCS even in the initial phases of smoking and tobacco addiction. Of note, our sample was composed of adolescents who were either just experimenting with smoking (average of seven cigarettes *per month*) or had never smoked. The effect size of this correlation was small, likely due to the limited variability in self-reported craving ratings, especially among nonsmokers, most of whom reported zero craving for cigarettes. Future studies in larger sample sizes and with more variability in smoking behavior could estimate the precise effect size for this type of between-subjects effect (Reddan et al., 2017, Han et al., 2022).

While some studies have reported no significant differences in self-reported craving between adolescent and adult smokers, they showed evidence for age- and dose-dependent effects of nicotine absorption on brain responses (Do and Galván, 2015, 2016; Rubinstein et al., 2011, Colyer-Patel et al., 2023). For example, adolescent smokers showed greater activation in the nucleus accumbens (NAcc) compared to adult smokers (Do and Galván, 2016), and light smoking adolescents did not show activity in certain regions typically involved in cue-reactivity in adult smokers studies (e.g. insula, thalamus, fusiform, temporal regions; Rubinstein et al., 2011). Possibly, activation patterns in some brain regions vary at later stages of addiction or as the brain achieves its development. Future cross-sectional and longitudinal studies could test the NCS across different age groups and during the development of smoking addiction to further characterize the sensitivity of the NCS across different age and smoking groups.

In line with the correlational findings, our study also showed that experimental smoking adolescents reported higher craving levels and NCS responses than non-smoking adolescents. This effect was driven by the relative difference between conditions, as the two groups did not differ in any of the conditions alone (Smoking>Baseline, Neutral>Baseline and Romantic>Baseline). This finding seems to support the sensitization-homeostasis theory, according to which symptoms of tobacco addiction appear even before the onset of regular smoking (DiFranza & Wellman, 2005). It is also in line with previous findings that craving is one of the first and most central symptoms of addiction (O'Brien et al., 1998; Gauld et al., 2023). As discussed above, even participants in the smoking group had light smoking habits and low-to-moderate ETS levels, which likely minimized group differences in smoking-related behavior.

Another key focus of this study was to assess how social context, such as exposure to second-hand smoke, might impact craving and craving-related brain activity. Results showed no associations between overall ETS exposure and cigarette craving or NCS responses to smoking compared to neutral and romantic social cues. However, compared to children, adolescents spend more time with peers than with their family and form more complex peer relationships (De Goede et al., 2009). Adolescents have an increased need for social approval by peers, which makes them more sensitive to peer acceptance versus rejection, compared to children or adults (Orben, Tomova & Blakemore, 2020). In contrast, the influence of adults and especially parents on adolescents' behavior may decrease during this age period (Knoll et al., 2015; Mercken et al., 2009). Indeed, exploratory analyses showed that greater relative smoking exposure from friends compared to family was associated with higher NCS responses to smoking compared to neutral and romantic social cues. This is in line with findings that people may adjust their behavior towards the observed norm of the in-group but away from the observed norm of an out-group (Izuma & Adolphs, 2013) and that different social contexts might play a critical role in shaping individuals' neural reactivity to smoking stimuli. Given the correlational nature of our analysis, it is also possible that experimental smoking adolescents are more likely to befriend other smoking rather than non-smoking adolescents, whereas their smoking behavior likely has little effects on parental ETS. We also note that peers' versus parents' smoking behavior may have different contributions at different stages of addiction (such as initiation versus ability to quit smoking later in life). Future studies should differentiate between social sources of environmental tobacco smoke exposure and further explore the social network dynamics of smoking initiation and cessation (Christakis & Fower, 2008) in adolescents. They could also use personalized fMRI stimuli or smoking cues featuring different groups of smokers, since personalized stimuli elicit stronger cravings compared to generic cues, and may be more effective in identifying an association between craving-related brain activity, ETS exposure and monthly smoking (Cho et al., 2008; Conklin et al., 2010).

Finally, we observed that the NCS responded most strongly to social romantic cues in adolescents, independently of the experimental group (Fig. 2a). This suggests that romantic social interactions may evoke neural craving responses similar to those for food and drugs in youth, consistent with the inherently rewarding nature of social stimuli. This result resonates with recent findings (Tomova et al., 2020) of shared midbrain responses to food cues after fasting and to social cues after social isolation. The group difference in NCS responses to smoking versus romantic cues reinforce our previous hypothesis, showing that experimental smokers might already be developing a relative neural preference for smoking-related rewards compared to alternative, social rewards. Despite the early stage of cigarette use in our participants, this pattern is consistent with the incentive sensitization theory of addiction, which argues that repeated drug use sensitizes the brain's reward circuits to drug cues, while diminishing responsiveness to natural rewards (Berridge & Robinson, 2016; Meyer, King & Ferrario, 2016) and especially social rewards (Heilig et al., 2016; Løseth et al., 2025). Future research should investigate the common and distinct neural bases of social, food, and drug cravings to better understand their interactions, especially during adolescence.

To our knowledge, this is the first study to address the association between current ETS exposure and craving-related brain activity using fMRI and an *a priori* defined brain signature of craving – the NCS. Using brain signatures integrates complex patterns of brain activity across multiple regions, offering more accurate and reliable insights into specific core processes such as craving (Woo et al., 2017, Han et al., 2022). Additionally, the longitudinal design and large sample size (N=148) allowed to track participants' smoking behavior over a year, providing insights into individual differences at early phases of smoking, useful to identify individuals at higher risk for developing nicotine dependence. As limitations, we note that the questionnaires employed only captured current ETS exposure and did not account for past exposure, which may also contribute to participants' behavioral and neural responses. Further, participants presented relatively low-to-moderate exposure to ETS and monthly smoking, which was relatively stable over time, which might explain the weak association between overall ETS exposure, monthly smoking, and brain functioning. Future studies could include participants with higher levels of ETS exposure and who are smoking daily, to cover the full spectrum of nicotine dependence in adolescence.

### *Conclusions*

Tobacco smoke is the first preventable cause of death worldwide. Understanding the early neural processes involved in tobacco craving and the development of addiction in vulnerable groups, such as adolescents, is therefore essential. This study examined how experimental tobacco use and exposure to environmental tobacco smoke (ETS) influence craving and craving-related brain activity in adolescents. Our findings revealed that NCS responses to smoking cues predicted higher self-reported craving and were higher compared to alternative social rewards in experimental smokers compared to nonsmokers. Additionally, both smoking behavior and NCS responses were shaped by the relative amount of ETS exposure from peers compared to exposure by parents and family members. Further studies should employ cohorts of adolescents with greater levels of smoking and ETS exposure, experimentally manipulate different types of social smoking cues and further investigate role of the context in which smoking exposure happens.

## Methods

### *Participants*

The data for this study was obtained from the MRI project on Nicotine Dependence in Adolescents, a longitudinal study of 148 adolescents aged between 14 and 19 from secondary schools and institutes around Nijmegen, in the Netherlands (Boormans et al., 2021; Dieleman et al., 2020; Dieleman et al., 2022). Non-smokers included 48 participants who never smoked a complete cigarette in their life ($M_{age}$= 16.56±1.09 years). Experimental smokers included 100 participants who smoked between 5 and 500 cigarettes in their lifetime ($M_{age}$=16.51±1.16 years) (Fig 1a). Participants were excluded from the study if they 1) were daily smokers, 2) took psychoactive medication that could not be interrupted for 24 hours, 3) had a history of physical or neurological disease/s, and/or 4) presented fMRI contraindications. All hypotheses, methods and analyses were preregistered on the Open Science Framework (OSF) website (https://osf.io/mc8v6/).

### *Measures*

*Sociodemographic factors*. Sociodemographic data (birth date, gender and age) was collected through an online questionnaire implemented in Qualtrics (https://www.qualtrics.com/). Pubertal status was determined using the Pubertal Development Scale (Petersen et al., 1988), evaluating stages of pubertal development with five questions on secondary sexual characteristics that are both gender neutral (body growth, pubic hair, skin changes, voice change), and gender specific (facial hair for boys and breast development and menarche for girls).

*Familial risk*. Familial risk assessed the participants' familiar vulnerability to develop nicotine dependence (Vink et al., 2005). A composite risk score was computed after completion of three questionnaires by one of the participant's parents. First, parents were asked to report whether they smoked at the moment of the experimental session, and their daily cigarette use. Answers divided them in four categories: nonsmoking (0), smoking less than 10 cigarettes per day (1), smoking between 10 and 19 cigarettes per day (2) and smoking 20 or more cigarettes per day (3). Second, the level of nicotine dependence during the period in which the parents smoked the most was measured by the Fagerström test for nicotine dependence (Heatherton et al., 1991), with higher scores indicating greater nicotine dependence symptoms. Third, the smoking behavior of their own parents, the participants' grandparents, was assessed. A score of 0 was assigned if answers were: "non-smokers", "ex-smokers", "smoke occasionally" and "smoke 1-10 cigarettes per day"; whereas a score of 1 was assigned if the answer was "smoke 10 or more cigarettes a day". Scores from both parents for the three questionnaires were summed and averaged, resulting in a total score ranging from 0 to 16.

*Smoking during pregnancy*. Participants' parents were asked: "Did you/your wife smoke during the pregnancy of your son/daughter?". Answers could be "yes" (1) or "no" (0).

*Alcohol and cannabis use*. Participants completed the Alcohol Use Disorder Identification Test (AUDIT, Saunders, Aasland & Babor, 1993). Its scores range from 0 to 40, with higher scores indicating more frequent and problematic alcohol use. Participants also reported whether they had ever used cannabis, and if yes, provided the number of occasions for lifetime cannabis use.

*ETS exposure*. Participants answered the question: "How often does X smoke when you are around?", where X referred to relatives (father, mother, siblings, stepmother, stepfather), friends (best friend and other friends) or "others" in the environment. Responses were rated on a scale from 0 to 8, ranging from "My X smokes, but when I am not around", to "More than five times a day". Participants' responses were combined for six people including relatives, friends and others, yielding a total ETS score ranging from 0 to 48. Higher values indicated greater ETS exposure.

*Monthly smoking*. The number of cigarettes smoked per month was used as the most reliable measure of smoking frequency in the group of experimental smoking adolescents (Dieleman et al., 2020; Dieleman et al., 2022). Participants indicated the answer that best described their smoking habits between 1: "I smoke at least once a day" and 9: "I never smoked". Participants who chose options 6, 7 or 8 quit smoking, so their monthly cigarette count was 0. Participants who chose option 2 reported their weekly cigarette consumption, which was converted to a monthly consumption (multiplied by 52 and divided by 12). Participants who selected options 3, 4 or 5 directly indicated their average monthly cigarette consumption.

*Cotinine levels*. Saliva samples were collected to measure cotinine levels and were analyzed with liquid chromatography and mass spectrometry (quantification limit of <1.0 µg/L). Cotinine is a metabolite of nicotine and is indicative of both ETS exposure and active use of cigarettes in the last 72 hours. Its measurement served as an objective and biological measurement of both recent active smoking and recent ETS exposure.

*Self-report measures related to smoking behavior*. Nicotine dependence was quantified using the Fagerström Test for Nicotine Dependence (FTND, Heatherton et al., 1991). The Hooked On Nicotine Checklist (HONC) was used to evaluate symptoms of nicotine dependence and quit attempts, even among non-smokers (DiFranza et al., 2004). The Autonomy Over Tobacco Scale (AOTS) was employed to gauge autonomy over smoking behavior and cue-induced craving (DiFranza et al., 2009). The Craving Smoking Item used a visual analog scale ranging from 0 to 100 to assess the intensity of current cravings for cigarettes. The Questionnaire for Smoking Urges (QSU) also measured participants' current cravings (Cox et al., 2001). The type of motivation to quit smoking was evaluated by asking participants to indicate whether they wanted (desire), ought (duty) or intended (intention) to stop smoking (Smit et al., 2011). Additionally, participants indicated which statement better described them between (1) “I plan to continue smoking and not cut down” and (9) “I plan to quit within 10 days”. Smoking-related self-efficacy (SSE) was evaluated through a questionnaire assessing individuals' confidence in refraining from smoking across various situations (Etter et al., 2000). The Smoking Specific Parenting (SSP) questionnaire assessed parenting practices discouraging children from smoking initiation (e.g. smoking agreement, house rules, frequency and quality of communication related to smoking) (Harakeh et al., 2005; Ennett et al., 2001).

### *Procedures and task*

During the fMRI scanning, participants were presented with a series of images that appeared for 3 to 5 seconds, one at a time, in the middle of a screen (Boormans et al., 2021; Dieleman et al., 2022, Fig.1b). They were instructed to pay attention to the images and to press a button as fast as possible when a new picture appeared on the screen. The task was composed of 24 blocks, with each block including four pictures of the same stimulus category (i.e., Smoking, Neutral, or Romantic). In total, 32 photographs of smoking-related interactions (people smoking or holding smoking-related objects), 32 photographs of non-smoking-related interactions (neutral social interactions) and 32 photographs of romantic interactions (a couple kissing, hugging, holding hands) were presented (Fig.1b, 1c). Each image was presented once during the whole task. The pictures in the three conditions were matched on visual properties (e.g. color, luminance, complexity), the amount of people, and their gender. Each block lasted 16s, and a fixation cross was displayed on the screen for 6.5 to 10 seconds to divide blocks. During the task, participants had twice a 15-second break.

### *fMRI data acquisition*

A Siemens 3 Tesla Skyra MRI scanner (Siemens Medical system, Erlangen, Germany) equipped with a 32-channel coil was used to acquire all MRI data. Multi-echo echoplanar imaging (EPI) enabled to acquire functional T2*-weighted imaging, capturing 39 axial slices in interleaved ascending order (voxel size 3.5x3.5x3.0 mm; matrix 64x64; repetition time (TR) 2020ms; echo times (TE) 7ms, 16.3ms, 26ms, 35ms, and 44ms; flip angle 80°).

### *Data analysis*

*fMRI data analysis*. Data preprocessing was performed using SPM12 (www.fil.ion. ucl.ac.uk/spm). Preprocessing steps included realignment, echo combination using the PAID method (Poser et al., 2006), coregistration, normalization into the MNI space, and smoothing with a full-width at half-maximum (FWHM) kernel of 8mm. An independent component analysis was performed using FSL to remove movement-related noise (www.fmrib.ox.ac.uk/fsl; ICA-AROMA) (Pruim et al., 2015).

The smoking cue-reactivity task was modeled using a general linear model (GLM) for each participant, with separate regressors for Smoking, Neutral and Romantic cues, and a regressor of no interest for the instructions. The smoking cue-reactivity blocks were modeled as 16-second boxcars, time-locked to the onset of the first image in each block. To eliminate low-frequency noise, a 128-second temporal high-pass filter was applied to the time series of the functional images. Group level analyses were performed using SPM12 and the CanlabCore toolbox (https://github.com/canlab/CanlabCore) on Matlab 2023a, with image visualization in MRIcroGL. These included a whole brain two-sample t-test to compare Experimental smoking and Non-smoking adolescents' activations for the [Smoking>Neutral] and [Smoking>Romantic] contrasts. Whole-brain T-maps generated from two-sample t-tests were overlaid on the mni152 brain template and were thresholded at an uncorrected voxel-level significance of $p<0.001$.

*NCS application*. Machine-learning-based patterns of brain activity – termed *neuromarkers* or *brain signatures* – have emerged as powerful tools for decoding mental states from neuroimaging data (Woo et al., 2017, Kragel et al., 2018). Based on distributed brain activity patterns, the Neurobiological Craving Signature (NCS) has recently been developed to predict cue-induced drug and food craving with high accuracy in different types of substance users (Koban et al., 2023). The NCS consists of a weight map that specifies each voxel's positive or negative contribution (i.e. statistical weight) to the prediction of craving, plus an intercept. We applied the NCS to the first-level contrast images for each of the three cue conditions (Smoking, Neutral, Romantic) versus implicit baseline, by computing the dot product between individual contrast images and the NCS (Fig.1c), yielding one scalar value per participant and condition.

*Statistical analyses.* The two groups were compared with two-sided Mann-Whitney U-tests or one-sample Wilcoxon tests (variance of 0) for continuous variables (age, pubertal status, familial risk, alcohol and cannabis use, cotinine levels, ETS exposure, monthly smoking, self-reported craving and questionnaire scores), and with chi-squared tests for dichotomous variables (gender and smoking behavior of the mother during pregnancy). Questionnaires scores were compared across time points (baseline, six-months and one-year follow-up) with repeated-measures ANOVAs corrected with the Greenhouse Geisser correction if sphericity was not respected for continuous variables and with Cochran's Q tests for dichotomous variables. To compare the NCS responses across groups and cue-types we performed a parametric mixed-effects ANOVA with a within-subjects factor (cues: Smoking, Neutral, Romantic) and a between-subjects factor (group: Non-Smokers and Experimental Smokers). Moreover, one-tailed Mann-Whitney U tests were used to compare the NCS responses to Smoking relative to Neutral and Romantic cues for Experimental smokers and Non-smokers. To test the association between individual differences in ETS exposure, monthly smoking, cotinine level, and self-reported craving on the one hand, and NCS response (difference between Smoking and Neutral/Romantic cues) on the other hand, we performed a Kendall's tau correlation analysis across all participants. Gender and age were not included as covariates as they had no effect on NCS responses (see Supplementary Fig. S3). To assess if the NCS activity at baseline was positively associated with smoking-related behavior at follow-up, and thus to identify at-risk individuals, we computed Kendall's tau correlation analyses between the NCS response at baseline (difference between Smoking > Neutral and Smoking > Romantic cues) and follow-up smoking-related behavior. We considered two principal outcomes of interest: 1) the level of self-reported craving and 2) monthly smoking. Significance level was set at 0.05 for all statistical tests.

*Deviations from preregistration.* In addition to comparing neural cue-reactivity (NCS) responses to social smoking cues versus neutral social interaction cues, we also examined responses to social smoking cues relative to alternative social reward cues (romantic images). This comparison was included to determine whether neural responses to social smoking cues reflected general sensitivity to socially rewarding stimuli or were specific to smoking-related cues.

We initially planned to test a mediation model with ETS exposure as the predictor, baseline NCS responses (Smoking vs Neutral) as the hypothetical mediator, and monthly smoking as the outcome variable. However, as suggested by the non-significant correlation between overall ETS exposure and monthly smoking ($\tau=0.10$, $p=0.11$, Fig. 3a), the direct path *c'* testing the relationship between ETS exposure and monthly smoking was not significant ($\beta=0.18$, STE=0.10, $t=1.82$, $p=0.07$). Therefore, the mediation model was not further tested.

Similarly, a preregistered PCA on the follow-up questionnaires did not reveal any significant relationships between baseline craving-related brain activity and future smoking-related behavior in Experimental smoking adolescents.

### Data availability

Deidentified data used in this study is available upon request and will be made available in a public repository upon publication of the manuscript.

### Code availability

Matlab code and R scripts used for analyses will be uploaded on https://github.com/ldmk.

### Acknowledgements

The study was supported by a VENI grant awarded to ML from the Netherlands Organisation for Scientific Research (NWO: Nederlandse Organisatie voor Wetenschappelijk Onderzoek, nr. 451-15-029) and a Starting Grant from the European Research Council (ERC-StG SOCIALCRAVING, 101041087) awarded to LK.

# Supplementary materials

## Responses of the Neurobiological Craving Signature to smoking versus alternative social rewards predict craving and monthly smoking in adolescents

Maddalena Tamellini[1], Joyce Dieleman[2,3,4], Guillaume Sescousse[1], Maartje Luijten[3], & Leonie Koban[1]*

[1] Lyon Neuroscience Research Center (CRNL), INSERM U1028, CNRS UMR5292, Claude Bernard Lyon 1 University, Lyon, France.

[2] Trimbos Institute, Utrecht, The Netherlands.

[3] Behavioral Science Institute, Radboud University, Nijmegen, The Netherlands.

[4] GGD Noord- en Oost Gelderland, Warnsveld, The Netherlands.

*contact: leonie.koban@cnrs.fr

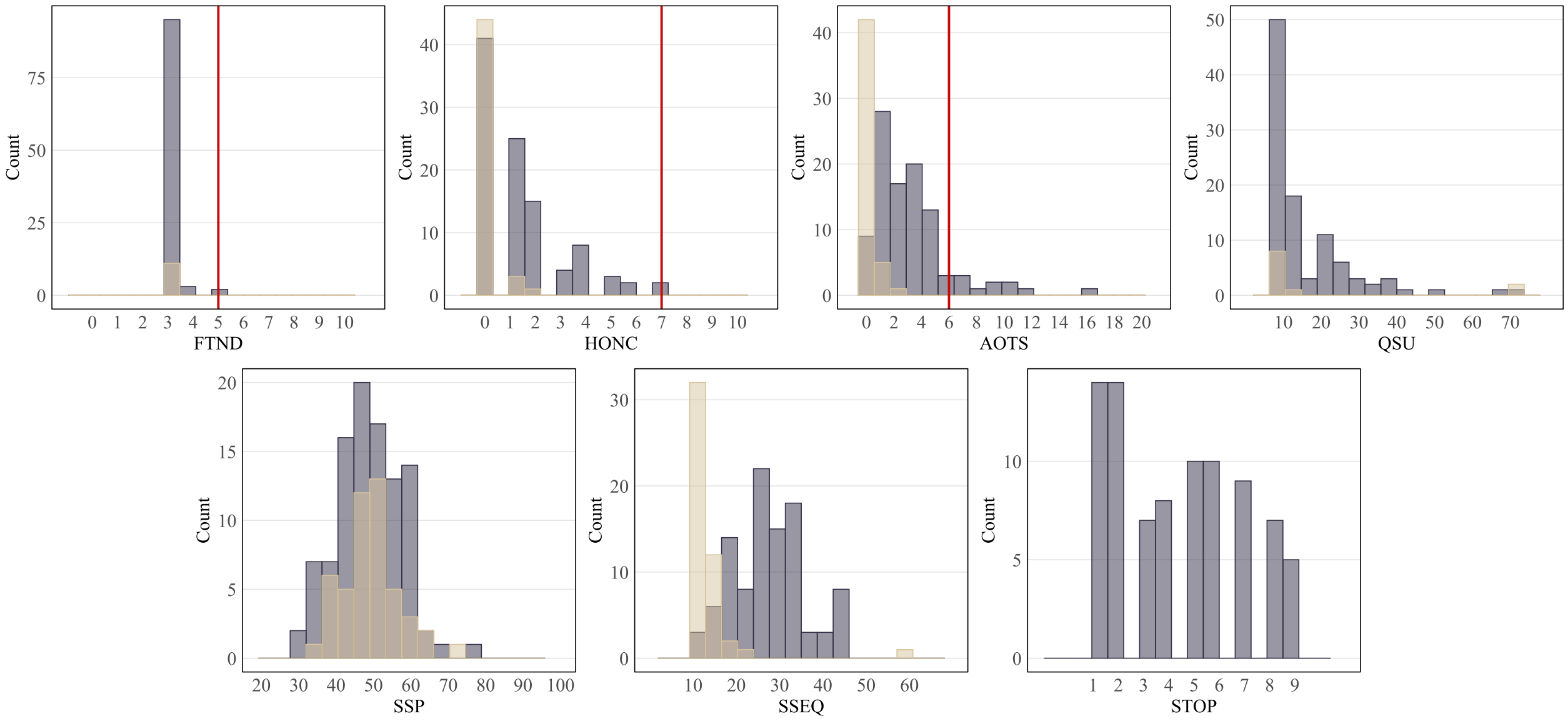

**<u>Figure S1:</u> Baseline questionnaires scores distribution for Experimental Smokers and Non-smokers.** Histograms representing the number of participants within each score range (Non-smokers in beige, Experimental smokers in dark blue). In red, the line indicating the conventional threshold for high dependence scores for the FTND, HONC and AOTS. Only 11 out of 48 Non-smokers answered the FTND and the QSU. Only 84 out of 100 Experimental smokers (and none of the Non-smokers) answered the STOP questionnaire. FTND, Fagerström Test for Nicotine Dependence; HONC, Hooked On Nicotine Checklist; AOTS, Autonomy Over Tobacco Scale; QSU, Questionnaire for Smoking Urges; SSEQ, Smoking-related Self-Efficacy; SSP, Smoking Specific Parenting; STOP, Willingness to stop smoking.

**Supplementary Table S1: Experimental smoking adolescents' smoking-behavior and questionnaire scores at baseline and follow-up timepoints.**

| *Questionnaire* | *Experimental smokers* | | | | | |
|---|---|---|---|---|---|---|
| | *Baseline* | *Six-months* | *One-year* | *N* | *F / Q* | *p* |
| **Monthly smoking*** | 6.37 ± 9.49 | 6.61 ± 11.83 | 12.34 ± 29.50 | 79 | 3.681 | 0.048 |
| Self-reported craving | 17.26 ± 20.86 | 17.08 ± 20.24 | 14.18 ± 17.87 | 97 | 1.634 | 0.199 |
| FTND | 3.05 ± 0.27 | 3.03 ± 0.32 | 3.14 ± 0.61 | 79 | 1.648 | 0.203 |
| **HONC*** | 1.40 ± 1.73 | 2.11 ± 2.07 | 1.95 ± 1.99 | 100 | 7.623 | $8.666 \times 10^{-4}$ |
| AOTS | 3.16 ± 2.91 | 3.44 ± 3.09 | 3.88 ± 3.57 | 97 | 2.348 | 0.098 |
| QSU | 16.61 ± 11.53 | 17.80 ± 11.16 | 17.59 ± 12.98 | 97 | 0.411 | 0.664 |
| SSEQ | 27.72 ± 8.44 | 27.60 ± 9.19 | 27.90 ± 10.18 | 97 | 0.057 | 0.945 |
| SSP | 48.95 ± 8.65 | 48.97 ± 9.72 | 47.16 ± 8.67 | 97 | 2.567 | 0.079 |
| Stop | 4.00 ± 2.43 | 4.02 ± 2.63 | 4.51 ± 2.66 | 57 | 0.934 | 0.396 |
| Duty *(% Yes)* | 21.05 | 28.07 | 35.09 | 57 | 3.429 | 0.180 |
| Desire *(% Yes)* | 21.05 | 28.07 | 33.33 | 57 | 2.552 | 0.279 |
| Intention *(% Yes)* | 22.81 | 31.58 | 35.09 | 57 | 3.000 | 0.223 |

*Note:* Continuous variables are reported as mean ± std. Dichotomous variables (duty, desire and intention to quit smoking) are reported as the proportion of participants that answered "yes" to the question "Do you need/want/plan to quit smoking?". Variables for which values significantly differed between timepoints are indicated in bold (repeated-measures ANOVAs for continuous variables using the Greenhouse Geisser correction if sphericity was not respected, Cochran's Q test for dichotomous variables, *p<0.05). Monthly smoking (cigarettes/month); Self-reported craving (0-100 scores on a Visual Analog Scale); FTND, Fagerström Test for Nicotine Dependence; HONC, Hooked On Nicotine Checklist; AOTS, Autonomy Over Tobacco Scale; QSU, Questionnaire for Smoking Urges; SSEQ, Smoking-related Self-Efficacy; SSP, Smoking Specific Parentings; Stop, Motivation to quit smoking; Desire, Desire to quit smoking; Duty, Duty to quit smoking; Intention, Intention to quit smoking.

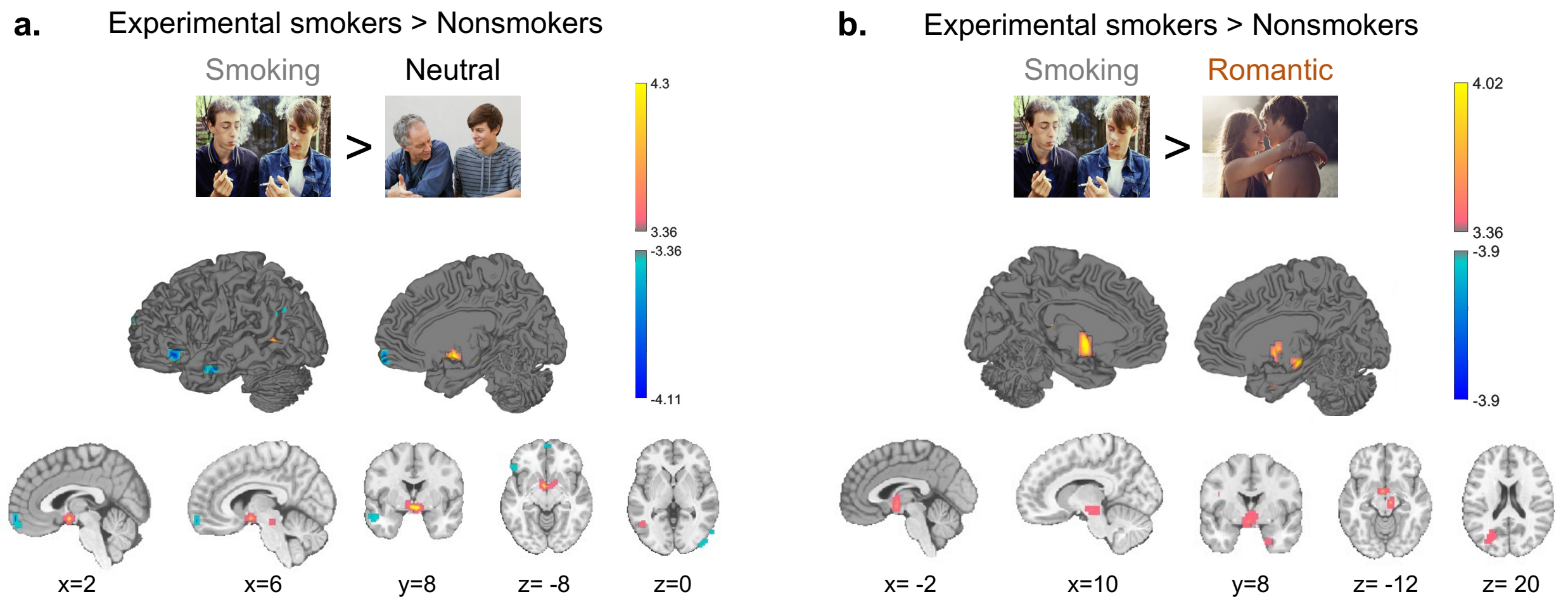

**Figure S2: Univariate differences in brain responses to [Smoking>Neutral] and [Smoking>Romantic] cues between Experimental smokers (N=100) and Non-smokers (N=48). a.** Whole-brain activations for the [Smoking>Neutral] contrast in Experimental smoking compared to Non-smoking adolescents showed an increased response in the ventral striatum, midbrain, visual cortex and in the fundus of the superior temporal area, and a decreased response in the ventrolateral prefrontal cortex, middle temporal gyrus and lateral occipital cortex (voxel-level uncorrected $p<0.001$). These regions did not survive cluster correction. **b.** Whole-brain activations for the [Smoking>Romantic] contrast in Experimental smoking compared to Non-smoking adolescents showed an increased activation in the ventral striatum, midbrain, hypothalamus, parahippocampal gyrus, ventrolateral prefrontal cortex and occipital cortex (voxel-level uncorrected $p<0.001$). These regions did not survive cluster correction.

**Supplementary table S2: [Smoking>Neutral] univariate activations in Experimental Smokers compared to Non-smokers.**

| **Positive effects** | | | | | | | |
|---|---|---|---|---|---|---|---|
| *Atlas region* | *Side* | *Label description* | *Volume (mm³)* | *X* | *Y* | *Z* | *maxZ* |
| Midb | R | Midbrain | 512 | 5 | -25 | -14 | 3.5993 |
| V3B | R | Area V3B, supplementary V3 | 528 | 29 | -63 | 21 | 3.4746 |
| IP0 | R | Area IntraParietal 0 | 384 | 29 | -60 | 32 | 3.4221 |
| FST | L | Fundus of the superior temporal area | 1248 | -48 | -60 | 4 | 3.8381 |
| Multiple regions | | Pallidum L, Olfactory R, Caudate R (Ventral Striatum) | 5136 | 1 | 4 | -11 | 4.9388 |
| **Negative effects** | | | | | | | |
| *Atlas region* | *Side* | *Label description* | *Volume (mm³)* | *X* | *Y* | *Z* | *maxZ* |
| 10v | R | Area 10v, cingulate ventral frontal | 1952 | 1 | 63 | -11 | -4.2655 |
| 10d | L | Area 10d, cingulate ventral frontal | 384 | -13 | 67 | 21 | -3.6855 |
| TE1a | L | Area TE1 anterior, temporal lateral | 1528 | -59 | 0 | -21 | -3.7706 |
| 47l | L | Area 47 lateral, cingulate ventral frontal | 1376 | -52 | 28 | -7 | -3.7221 |
| PFm | L | Parietal area F part m, parietal inferior lobule | 648 | -59 | -67 | 28 | -3.5094 |
| LO1 | R | Area lateral occipital 1, visual MT+ | 2160 | 43 | -88 | 4 | -4.8408 |
| No label | | | 384 | 57 | -70 | 0 | -3.7003 |

**Supplementary table S3: [Smoking>Romantic] univariate activations in Experimental Smokers compared to Non-smokers.**

| **Positive effects** | | | | | | | |
|---|---|---|---|---|---|---|---|
| *Atlas region* | *Side* | *Label description* | *Volume (mm³)* | *X* | *Y* | *Z* | *maxZ* |
| Midb | R | Midbrain (Ventral Striatum) | 2208 | 8 | -21 | -11 | 4.2483 |
| IP0 | L | Area intraparietal 0, parietal inferior lobule | 5112 | -27 | -70 | 25 | 3.8346 |
| IFJp | L | Area IF Jp, cingulate vlPFC | 288 | -45 | 0 | 25 | 3.4572 |
| PeEc | R | Perirhinal Ectorhinal Cortex | 672 | 19 | 0 | -35 | 3.7083 |
| Thal_Hythal | | Hypothalamus | 4232 | -3 | 0 | -7 | 4.3083 |
| **Negative effects** | | | | | | | |
| *Atlas region* | *Side* | *Label description* | *Volume (mm³)* | *X* | *Y* | *Z* | *maxZ* |
| TPOJ2 | R | Area TemporoParietoOccipital Junction 2 | 384 | 64 | -60 | 11 | -3.3626 |
| No label | | | 288 | 61 | -67 | 14 | -3.9033 |

*Note:* Positive and negative effects for the contrast [Smoking>Neutral] (Table S2) and [Smoking>Romantic] (Table S3) (uncorrected p<0.001 across the whole-brain gray matter mask). Atlas regions are labeled based on a combination of parcellations available on GitHub: https://github.com/canlab/Neuroimaging_Pattern_Masks/tree/master/Atlases_and_parcellations/2018_Wager_combined_atlas (R: Right, L: Left). This repository includes multiple atlases and other meta-analytic and multivariate maps. Tools for manipulating and analyzing this and other atlases are in the CANlab Core Tools repository: https://github.com/canlab/CanlabCore.

**Supplementary table S4: Association between the NCS responses to Smoking versus Neutral and Romantic cues and smoking-related behavior at follow-up.**

| *Variables* | | Six-months follow-up | | One-year follow-up | | Baseline | |
|---|---|---|---|---|---|---|---|
| | | Monthly smoking | Self-reported craving | Monthly smoking | Self-reported craving | NCS Smoking>Neutral | NCS Smoking>Romantic |
| Six-months follow-up | Monthly smoking | / | $1.848 \times 10^{-2}$ | $1.166 \times 10^{-7}$ | $5.326 \times 10^{-3}$ | 0.947 | 0.696 |
| | Self-reported craving | 0.185 | / | 0.197 | $3.952 \times 10^{-8}$ | 0.826 | 0.735 |
| One-year follow-up | Monthly smoking | 0.451 | 0.105 | / | $5.035 \times 10^{-8}$ | 0.483 | 0.057 |
| | Self-reported craving | 0.224 | 0.421 | 0.454 | / | 0.722 | 0.372 |
| Baseline | NCS Smoking>Neutral | -0.005 | 0.016 | -0.054 | 0.026 | / | $2.270 \times 10^{-4}$ |
| | NCS Smoking>Romantic | 0.029 | 0.024 | 0.148 | 0.065 | 0.255 | / |

*Note.* Correlation matrix between Experimental smokers' baseline NCS responses to [Smoking>Neutral] and [Smoking>Romantic] cues and follow-up monthly smoking (cigarettes/month) and self-reported craving. Kendall's tau pairwise correlations were used to test these associations. Kendall's tau correlation coefficients are indicated below the diagonal and p-values are indicated above the diagonal.

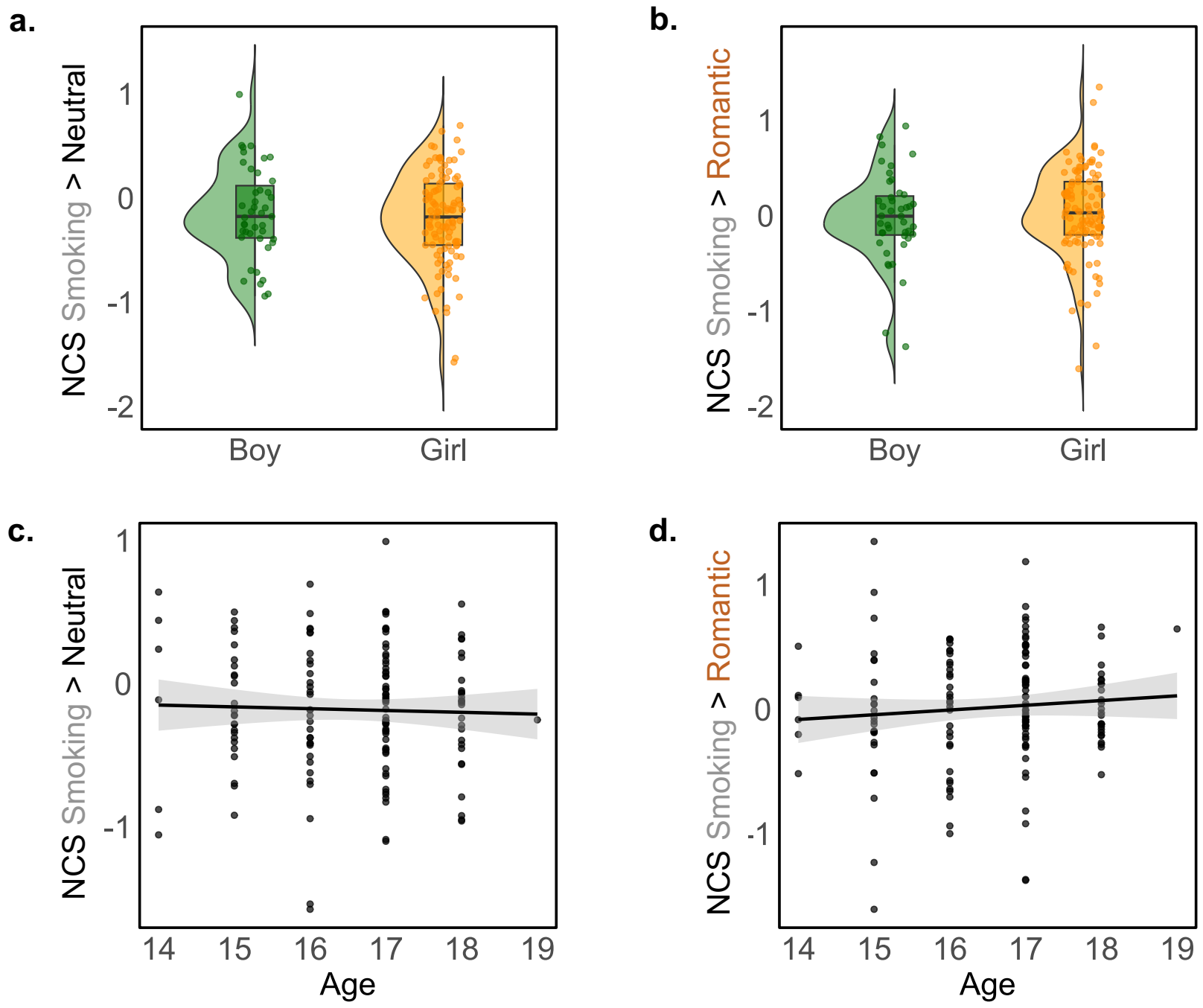

**Figure S3:** **Effects of gender and age on NCS responses to Smoking versus Neutral and Romantic cues. a.** Raincloud plot showing the absence of a gender effect on NCS responses to [Smoking > Neutral] cues (Boys: N=43, green; Girls: N=105, orange. Two-sided Mann-Whitney U-test U=2106.5, p=0.525) and **b.** NCS responses for [Smoking > Romantic] cues (Boys: N=43; green, Girls: N=105, orange. Two-sided Mann-Whitney U-test, U=2350.5, p=0.696). **c.** Absence of correlation between age and NCS responses to [Smoking > Neutral cues] (Kendall's tau correlation, N=148, $\tau$=0.05, p=0.38) and **d.** [Smoking > Romantic] cues (Kendall's tau correlation, N=148, $\tau$= -0.03, p=0.64).

**Supplementary discussion:**

Our exploratory mass-univariate findings paralleled and complemented the NCS findings. Experimental smokers exhibited increased responses in the midbrain compared to Non-smoking adolescents for Smoking-related versus both Neutral and Romantic cues. The midbrain, including the ventral tegmental area and the substantia nigra, is part of the dopaminergic mesocorticolimbic system, which plays an important role in reward, motivation, and addiction (Hyman, Malenka & Nestler, 2006). The increased midbrain response to Smoking cues may reflect increased dopaminergic response, suggesting a stronger incentive salience and craving triggered by these cues in Experimental smokers (Due et al., 2002; Volkow, Fowler & Wang, 2003). Additionally, the reduced activation of ventrolateral prefrontal regions in experimental smokers might also reflect reduced cognitive regulation of craving (Kober et al., 2010). These findings overall suggest that even early and limited smoking experiences may alter fundamental reward- and cognitive control-related mechanisms, potentially contributing to the reinforcement of smoking behavior. Alternatively, given our quasi-experimental experimental design, they may also reflect a pre-existing higher vulnerability of the experimental smokers compared to the never-smokers. Future research could aim to include non-smoking participants at high risk of developing nicotine dependence in the future, or participants with greater monthly cigarette consumption and higher exposure to environmental tobacco smoke.

**Supplementary references:**